\documentclass[a4paper,11pt]{article}
\usepackage{pos}
\usepackage{braket}
\title{Astrophysical neutrino oscillations accounting for neutrino charge radii }

\author[a]{Konstantin Kouzakov}
\author[a]{Fedor Lazarev}
\author[a]{Vadim Shakhov}
\author*[a]{Konstantin Stankevich}
\author[a,b]{Alexander Studenikin}

\affiliation[a]{ Faculty of Physics, Lomonosov Moscow State University,\\ Moscow 119991, Russia}

\affiliation[b] {Joint Institute for Nuclear Research,\\ Dubna 141980, Moscow Region, Russia}

\emailAdd{kl.stankevich@physics.msu.ru}
\emailAdd{studenik@srd.sinp.msu.ru}

\abstract{We derive for the first time an effective neutrino evolution Hamiltonian accounting for neutrino interactions with external magnetic field due to neutrino charge radii and anapole moment. The results are interesting for possible applications in astrophysics. 
	}

\FullConference{%
  40th International Conference on High Energy physics - ICHEP2020\\
  July 28 - August 6, 2020\\
  Prague, Czech Republic (virtual meeting)
}

\begin{document}
\maketitle

It is well known that neutrino electromagnetic interactions \cite{Giunti:2014ixa} are important for neutrino evolution and oscillations in different astrophysical environments (see, for example, \cite{deGouvea:2012hg,deGouvea:2013zp,Abbar:2020ggq}). We consider for the first time neutrino flavour, spin and spin-flavour oscillations engendered by neutrino interactions with an external magnetic field due to neutrino charge radii and anapole moment. Note, that in this case only the toroidal and poloidal magnetic fields matters. We perform similar calculations that were performed for derivations of the neutrino effective evolution Hamiltonians in the presence of magnetic fields and moving matter \cite{Fabbricatore:2016nec,Pustoshny:2018jxb}. 

The neutrino electromagnetic interactions is described by the effective interaction Hamiltonian\cite{Giunti:2014ixa} 

\begin{equation}
	H_{int} (x) = j_{\mu}^{(\nu)} (x) A^\mu (x) = \sum_{k,j}^{2} \bar{\nu}_k (x) \Lambda^\mu_{kj} \nu_j (x) A_\mu (x)
	,
\end{equation}
where $\Lambda^\mu_{kj}$ is the neutrino electromagnetic  vertex function. Here below, we are interested only in the charge radii $\dfrac{\braket{r^2}}{6}$ and anapole form factors $f^A$ in $\Lambda^\mu_{kj}$, thus we use

\begin{equation}
	\Lambda^\mu_{fi} (q) = (q^2 \gamma_\mu - q^\mu \gamma^\nu q^\nu) \left[\dfrac{\braket{r^2}_{fi}}{6} + f^A_{fi} \gamma_5 \right]
	.
\end{equation} 
Within calculations similar to those of \cite{Fabbricatore:2016nec,Pustoshny:2018jxb}, we get the following effective interaction Hamiltonian for the considered case:
 
\begin{equation}\label{Hamilt}
	H_{\alpha \alpha'}^{s s'} =  u_{s \alpha}^\dagger \left\{ \left[ \text{curl}\boldsymbol{B}\right]_{||} \left( \dfrac{\braket{r^2}_{\alpha \alpha'}}{6} + f^A_{\alpha \alpha'} \sigma_3 \right) + \left[ \text{curl}\boldsymbol{B}\right]_\perp \left( \gamma_{\alpha \alpha'}^{-1} f^A_{\alpha \alpha'} \sigma_1 -i \tilde{\gamma}_{\alpha \alpha'}^{-1} \dfrac{\braket{r^2}_{\alpha \alpha'}}{6} \sigma_2  \right) \right\} u_{s' \alpha'}
	,
\end{equation}
where $\left[ \text{curl}\boldsymbol{B}\right]_{||}$ is the component of the curl of the magnetic field parallel to the neutrino propagation and $\left[ \text{curl}\boldsymbol{B}\right]_\perp$ is the perpendicular component, $u_{s\alpha}$ is the neutrino spinor. The gamma factors are given by $\gamma_{\alpha}^{-1} =\frac{m_{\alpha}}{E_{\alpha}} ,
		\gamma_{\alpha\beta}^{-1} =\frac12\left(\gamma_{\alpha}^{-1} + \gamma_{\beta}^{-1}\right) ,
		\tilde \gamma_{\alpha\beta}^{-1} =\frac12\left(\gamma_{\alpha}^{-1} - \gamma_{\beta}^{-1}\right) .$

One can see that  $\left[ \text{curl}\boldsymbol{B}\right]_{||}$ is responsible for the flavour oscillations and $\left[ \text{curl}\boldsymbol{B}\right]_\perp$ is responsible for the spin and spin-flavour oscillations. Note, the spin and spin-flavour oscillations are suppressed by gamma factors. 
In the flavour basis $\nu_f = (\nu^L_e, \nu^L_x, \nu^R_e, \nu^R_x)$ the evolution Hamiltonian can be decomposed into two parts \begin{equation}
H^f = [\text{curl}\boldsymbol{B}]_{||}H^f_{1}+[\text{curl}\boldsymbol{B}]_{\perp}H^f_{2}
,
\end{equation}
where

\begin{equation}
H^{f}_1=
\begin{pmatrix}
\frac{\langle r^2 \rangle_{ee}}{6} + f^A_{ee} & \frac{\langle r^2 \rangle_{ex}}{6} + f^A_{ex} & 0  &  0 \\
\frac{\langle r^2 \rangle_{ex}}{6} + f^A_{ex} & \frac{\langle r^2 \rangle_{xx}}{6} + f^A_{xx} & 0 & 0 \\
0 & 0 & \frac{\langle r^2 \rangle_{ee}}{6} - f^A_{ee} & \frac{\langle r^2 \rangle_{ex}}{6} - f^A_{ex} \\
0 & 0 & \frac{\langle r^2 \rangle_{ex}}{6} - f^A_{ex} & \frac{\langle r^2 \rangle_{xx}}{6} - f^A_{xx}
\end{pmatrix},
\end{equation}

\begin{equation}
H^{f}_2=
\begin{pmatrix}
0 & 0 & \left(\frac{\braket{r^2}}{\gamma}\right)_{ee} + \left(\frac{f^A}{\gamma}\right)_{ee}  &  \left(\frac{\braket{r^2}}{\gamma}\right)_{ex} + \left(\frac{f^A}{\gamma}\right)_{ex} \\
0 & 0 & \left(\frac{\braket{r^2}}{\gamma}\right)_{e\mu} + \left(\frac{f^A}{\gamma}\right)_{e\mu} & \left(\frac{\braket{r^2}}{\gamma}\right)_{xx} + \left(\frac{f^A}{\gamma}\right)_{xx} \\
\left(\frac{f^A}{\gamma}\right)_{ee} - \left(\frac{\braket{r^2}}{\gamma}\right)_{ee} & \left(\frac{f^A}{\gamma}\right)_{ex} - \left(\frac{\braket{r^2}}{\gamma}\right)_{ex} & 0 & 0 \\
\left(\frac{f^A}{\gamma}\right)_{ex}-\left(\frac{\braket{r^2}}{\gamma}\right)_{ex} & \left(\frac{f^A}{\gamma}\right)_{xx} -\left(\frac{\braket{r^2}}{\gamma}\right)_{xx} & 0 & 0
\end{pmatrix}.
\end{equation}
The form factors in the flavour basis are defined as

\begin{equation}
	\begin{aligned}
		\langle r^2 \rangle_{ee} &= \langle r^2 \rangle_{11}\cos^2\theta + \langle r^2 \rangle_{22} \sin^2\theta + \langle r^2 \rangle_{12} \sin2\theta, \ \ \ \ \ 		f^A_{ee} = f^A_{11}\cos^2\theta + f^A_{22} \sin^2\theta +f^A_{12} \sin2\theta, \\
		\langle r^2 \rangle_{xx} &= \langle r^2 \rangle_{11}\sin^2\theta + \langle r^2 \rangle_{22}\cos^2\theta - \langle r^2 \rangle_{12}\sin2\theta, \ \ \ \
		 \ 	f^A_{xx} = f^A_{11}\sin^2\theta + f^A_{22}\cos^2\theta - f^A_{12}\sin2\theta,\\
		\langle r^2 \rangle_{ex} &= \langle r^2 \rangle_{12} \cos2\theta + \dfrac12 \left( \langle r^2 \rangle_{22} - \langle r^2 \rangle_{11} \right)\sin2\theta, \ \ \ \ \ \ \ f^A_{ex} = f^A_{12} \cos2\theta + \dfrac12 \left( f^A_{22} - f^A_{11} \right)\sin2\theta, \\
	\end{aligned}
\end{equation}
and

\begin{equation}
\begin{aligned}
\left(\dfrac{f^A}{\gamma}\right)_{ee} &= \dfrac{f^A_{11}}{\gamma_{11}} \cos^2\theta + \dfrac{f^A_{22}}{\gamma_{22}} \sin^2 \theta + \dfrac{f^A_{12}}{\gamma_{12}} \sin 2 \theta ,\ \ \ \ \ \ \ \left(\frac{\braket{r^2}}{\gamma}\right)_{ee} = \tilde{\gamma}_{12}^{-1} \frac{\braket{r^2}_{12}}{6} \sin 2 \theta , \\
\left(\dfrac{f^A}{\gamma}\right)_{xx} &= \dfrac{f^A_{11}}{\gamma_{11}} \sin^2\theta + \dfrac{f^A_{22}}{\gamma_{22}} \cos^2 \theta - \dfrac{f^A_{12}}{\gamma_{12}} \sin 2 \theta ,\ \ \ \ \ \ \ \left(\frac{\braket{r^2}}{\gamma}\right)_{xx} = - \tilde{\gamma}_{12}^{-1} \frac{\braket{r^2}_{12}}{6} \sin 2 \theta , \\
\left(\dfrac{f^A}{\gamma}\right)_{ex} &=\dfrac{f^A_{12}}{\gamma_{12}} \cos 2 \theta + \dfrac 1 2 \left( \dfrac{f^A_{22}}{\gamma_{22}} - \dfrac{f^A_{11}}{\gamma_{11}} \right) \sin 2 \theta , \ \ \ \ \ \ \ \ \left(\frac{\braket{r^2}}{\gamma}\right)_{ex} = \tilde{\gamma}_{12}^{-1} \frac{\braket{r^2}_{12}}{6} \cos 2 \theta . \\
\end{aligned} 
\end{equation}
The obtained evolution Hamiltonian can be used for the analysis of the flavour, spin and spin-flavour oscillations and corresponding resonances due to the neutrino electromagnetic interactions with the external magnetic field engendered by neutrino charge radii and anapole moment. This work was supported by the Russian Foundation for Basic Research under Grant No. 20-52-53022-GFEN-a. The work of KS was also supported by the Russian Foundation for Basic Research under Grant No. 20-32-90107.

\end{document}